\documentclass[aps,prb,twocolumn,groupedaddress,showpacs,floatfix]{revtex4-1}
\pdfoutput=1

\usepackage{graphicx}
\usepackage{dcolumn}
\usepackage{bm}
\usepackage{amsmath}
\DeclareMathOperator{\Tr}{Tr}

\begin{document}

\title{Topological effects and particle-physics analogies beyond the massless Dirac-Weyl fermion in graphene 
nanorings}

\author{Igor Romanovsky}
\email{Igor.Romanovsky@physics.gatech.edu}
\author{Constantine Yannouleas}
\email{Constantine.Yannouleas@physics.gatech.edu}
\author{Uzi Landman}
\email{Uzi.Landman@physics.gatech.edu}

\affiliation{School of Physics, Georgia Institute of Technology,
             Atlanta, Georgia 30332-0430}

\date{30 December 2012; Phys. Rev. B {\bf 87}, 165431 (2013)}

\begin{abstract}
Armchair and zigzag edge terminations in planar hexagonal and trigonal graphene nanorings are shown to 
underlie one-dimensional topological states associated with distinctive energy gaps and patterns 
(e.g., linear  dispersion of the energy of an hexagonal ring with an armchair termination versus parabolic 
dispersion for a zigzag terminated one) in the bands of the tight-binding 
spectra as a function of the magnetic field. A relativistic Dirac-Kronig-Penney model analysis of the 
tight-binding Aharonov-Bohm behavior reveals that the graphene quasiparticle in an armchair hexagonal ring 
is a condensed-matter realization of an ultrarelativistic fermion with a position-dependent mass term, akin 
to the zero-energy fermionic solitons with fractional charge familiar from quantum field theory and from 
the theory of polyacetylene. The topological origins of the above behavior are highlighted by contrasting it
with the case of a trigonal armchair ring, where we find that the quasiparticle excitations behave as 
familiar Dirac fermions with a  constant mass. Furthermore, the spectra of a zigzag
hexagonal ring correspond to the low-kinetic-energy nonrelativistic regime of a leptonlike massive fermion. 
A onedimensional relativistic Lagrangian formalism coupling a fermionic and a scalar bosonic field via a 
Yukawa interaction, in conjunction with the breaking of the $Z_2$ reflectional symmetry of the scalar field,
is shown to unify the above dissimilar behaviors.
\end{abstract}

\pacs{73.22.Pr, 73.21.-b, 11.10.-z, 73.23.Ra}   

\maketitle

\section{Introduction}
\label{secint}

Since its inception, relativistic quantum mechanics has been associated mainly with the fields of 
particle and high-energy physics. \cite{dira28,mcke87,bjorbook,grifbook} 
Recently, however, a {\it tabletop\/} version of relativistic quantum physics emerged, following the 
experimental isolation of graphene, which is a single-layer, planar honeycomb lattice of carbon atoms. 
\cite{geim04} Indeed, the two-dimensional (2D) quasiparticle excitations of neutral graphene near the 
Fermi level behave \cite{wall47,geim09} as massless neutrinolike fermions described by the celebrated 
Dirac-Weyl \cite{weyl29} (DW) equation. The scientific and technological potential for exploiting 
charge carriers and quasiparticles with relativistic behavior in tunable condensed-matter and atomic-physics 
systems is attracting much attention. \cite{zhu07,java08,chen10,mano12,zhan12}  
In this context, an important question, as yet only partly explored,
remains whether quasi-onedimensional (1D) graphene systems support exclusively DW massless or constant-mass 
Dirac fermions, or they can induce relativistic-quantum-field (RQF) behaviors that require 
consideration of position-dependent mass terms, reflecting generalized underlying bosonic scalar fields. 

In this paper, we show that planar graphene nanorings do indeed exhibit a rich variety of physics, ranging 
from sophisticated RQF regimes to more familiar cases of constant-mass fermions (both in the relativistic 
and nonrelativistic regimes). The emergence of these physical regimes depends on the specific combination of
topological factors associated with modifications of the graphene lattice, such as the type of edge termination
(i.e., armchair or zigzag) and the shape (i.e., hexagonal or trigonal) of the graphene ring.

To this end, we investigate the properties of the corresponding 
tight-binding \cite{geim09,roma12,roma12.2} (TB) spectra and of the associated Aharonov-Bohm \cite{ahar59} 
(the AB effect, which is a hallmark topological effect in condensed matter systems) oscillations of the 
magnetization, as a function of the magnetic field $B$. We then analyze the spectra and AB
characteristics with the help of a Dirac-Kronig-Penney (DKP) superlattice model, \cite{mcke87,krpe31} in the
spirit of the virtual ``band-structure'' model for the nonrelativistic AB effect. \cite{imry83} We find 
that the relativistic behavior in armchair rings requires a profound modification of the 1D Dirac equation 
\cite{dira28} through the introduction of a position-dependent mass term, in analogy with the 
fractional-charge, zero-energy topological modes in quantum field theory and in the theory of 
{\it trans\/}-polyacetylene. \cite{jack12,jack81,heeg88,camp01} In contrast, the zigzag-ring spectra may
correspond to the low-energy nonrelativistic regime of a leptonlike \cite{leptbook} massive particle, 
heavier than the electron. 

Planar graphene rings (and the associated AB spectra) have been recently investigated by a number of groups 
using tight-binding \cite{roma12.2,rech07,baha09,luo09,fert10.2,ma10} methods (for polygonal shapes with 
armchair or zigzag terminations), as well as continuum DW \cite{rech07,aber08,peet10} equations supplemented 
with infinite-mass boundary conditions (for idealized circular shapes). These earlier studies did not 
address the question of possible analogies to 1D quantum field theoretical models and particle physics. We 
stress that a prerequisite to raising and answering this question is the introduction by us of the virtual 
DKP superlattice model for the AB effect. For a review on recent experimental studies of the Aharonov-Bohm 
effect in graphene nanorings, see Ref.\ \onlinecite{trau12}.

In addition to planar graphene rings, graphene nanoribbons (GNRs) are another class of related quasi-1D 
systems. GNRs have attracted substantially more attention than graphene rings and their study gave rise to a 
vast body of theoretical \cite{tbgnr,dftgnr,dwgnr,revgnr} and experimental \cite{exgnr} literature. For 
graphene nanoribbons, it was found that a gap $\Delta_0$ may open at the Fermi energy, leading to an apparent 
analogy with the constant-mass Dirac fermion [see Ref.\ \onlinecite{tbgnr}(c)]. As elaborated below, 
for an armchair nanoribbon (aGNR), this gap arises from the topology of the armchair edge which mimicks the 
dimerized domains (i.e., formation of Kekul{\' e} unequal carbon bonds) in {\it trans\/}-polyacetylene. 
\cite{jack12,jack81,heeg88,camp01} Then in analogy with the scalar $Z_2$ kink-soliton associated with the
Peierls transition in {\it trans\/}-polyacetylene (or equivalently with the $Z_2$ kink-soliton used in the 
Jackiw-Rebbi fermionic RQF model \cite{jr76}) the qualitative features of the (fermionic) AB spectra of 
armchair graphene rings can be understood as resulting from an alternation (or lack of it) of two degenerate
dimerized domains associated with the arms of the graphene ring. We stress that the effective dimerization 
in the armchair GNRs and armchair graphene nanorings has a topological origin imposed by the
presence of the armchair edges, while the dimerization in {\it trans\/}-polyacetylene is due to the Peierls 
instability.\cite{heeg88,note4} These two different underlying processes, however, lead to similar 
results that are characterized by the breaking of the 1D $Z_2$ reflectional symmetry (see in particular
Secs.\ \ref{secdkphex} and \ref{seclag} below).

Our findings of quasiparticles in graphene with general position-dependent, and/or constant (rest), masses 
(unlike the massless neutrinolike quasiparticle in 2D graphene) is particularly interesting in light of 
increasing current interest in mass acquisition mechanisms, e.g., the Higgs mechanism in elementary 
particles \cite{brou64,higg64,gura64} and condensed-matter physics. \cite{ande62,sato11}

The predicted unprecedented emergent unfolding of fundamentally distinct physical regimes, namely
complex quantum-field theoretical ones versus nonrelativistic constant-mass ones, depending solely on the 
materials' shape and edge termination is to date unique to graphene. It will be of great interest to test 
signatures of such regime-crossover experimentally for specifically prepared graphene systems with 
atomic precision, \cite{exgnr} as well as to explore possible occurrence of such topological-in-origin 
physical behavior in other designer-Dirac-fermion artificial systems \cite{zhu07,mano12} or nanopatterned
artificial graphene.\cite{pell09}

The plan of the paper is as follows.

Sec. \ref{sectb} describes the Aharonov-Bohm tight-binding spectra for three characteristic planar graphene 
nanorings, i.e., an armchair hexagonal ring, an armchair trigonal ring, and a zigzag hexagonal ring.

Sec. \ref{secdkp} introduces the theoretical aspects of a relativistic 1D Dirac-Kronig-Penney model,  
based on the generalized Dirac equation. The DKP model describes the virtual superlattice 
associated with the Aharonov-Bohm effect.

Sec.\ \ref{secintdkp} presents the DKP interpretation of the tight-binding spectra calculated in
Sec.\ \ref{sectb}. The quasiparticle excitations in graphene nanorings are shown to exhibit
behavior associated with quantum field theoretical models for elementary particles beyond the massless
Dirac-Weyl fermion.   

Sec.\ \ref{seclag} discusses the full relativistic quantum field Lagrangian formalism that underlies the DKP
interpretation elaborated in Sec.\ \ref{secintdkp}. The Lagrangian formalism shows that the physics of 
quasiparticle excitations in planar graphene nanorings relates to mass acquisition and formation of 
fermionic solitons.   

Finally, Sec.\ \ref{seccon} presents our conclusions. 

\begin{figure*}[t]
\centering\includegraphics[width=13.5cm]{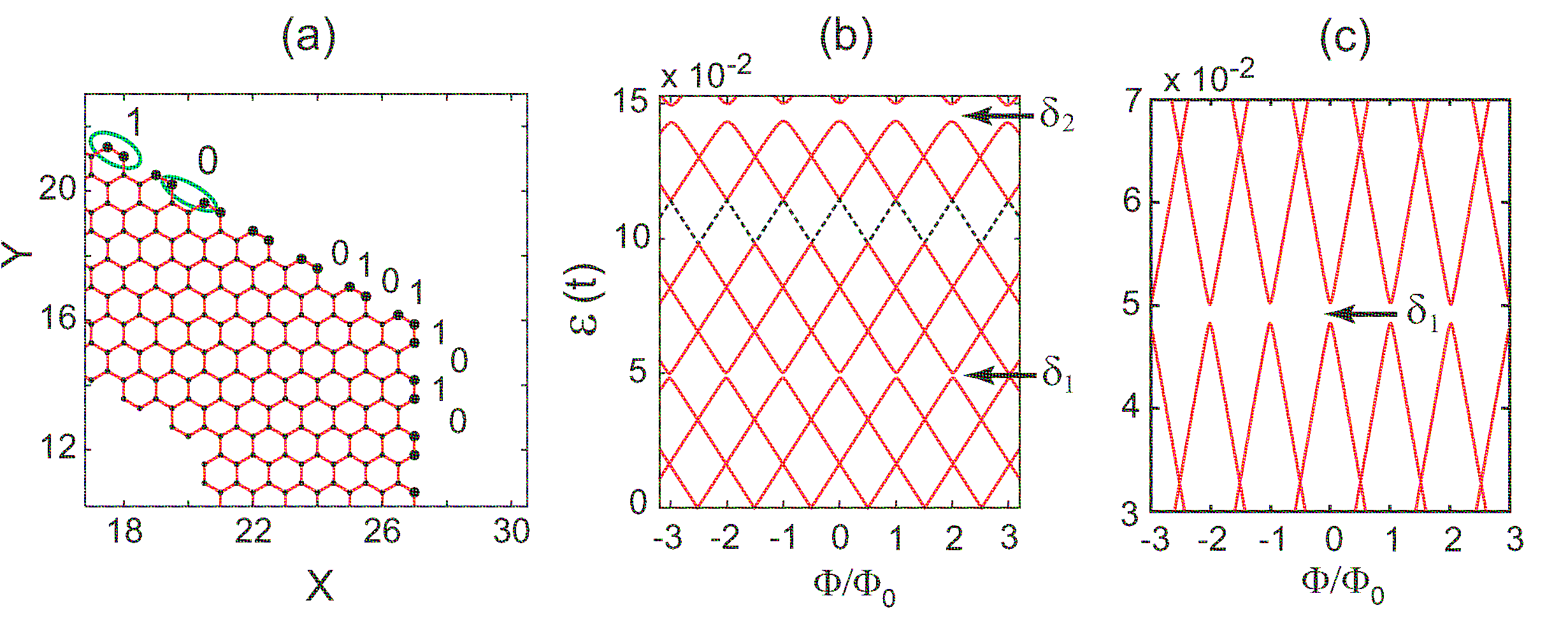}
\caption{
(Color online) 
   (a) Part of the hexagonal graphene ring (2718 carbon atoms) with {\it armchair\/} edges.
       1 and 0 denote the short and long carbon dimers, respectively.
       Lengths in units of the graphene lattice constant $a_0=0.246$ nm.   
   (b) TB single-particle spectrum as a function of the magnetic flux (magnetic field).
       Energies in units of the TB hopping-parameter $t=2.7$ eV. The dashed black line denotes the
       Fermi level for $N=14$ electrons. The arrows highlight the band gaps. 
   (c) Magnification of the TB spectrum around the $\delta_1$ band gap. $\delta_1 \sim 80$ K, 
       and thus it is easily detectable experimentally.
}
\label{tb_hex_arm1}
\end{figure*}

\begin{figure}[t]
\centering\includegraphics[width=5.5cm]{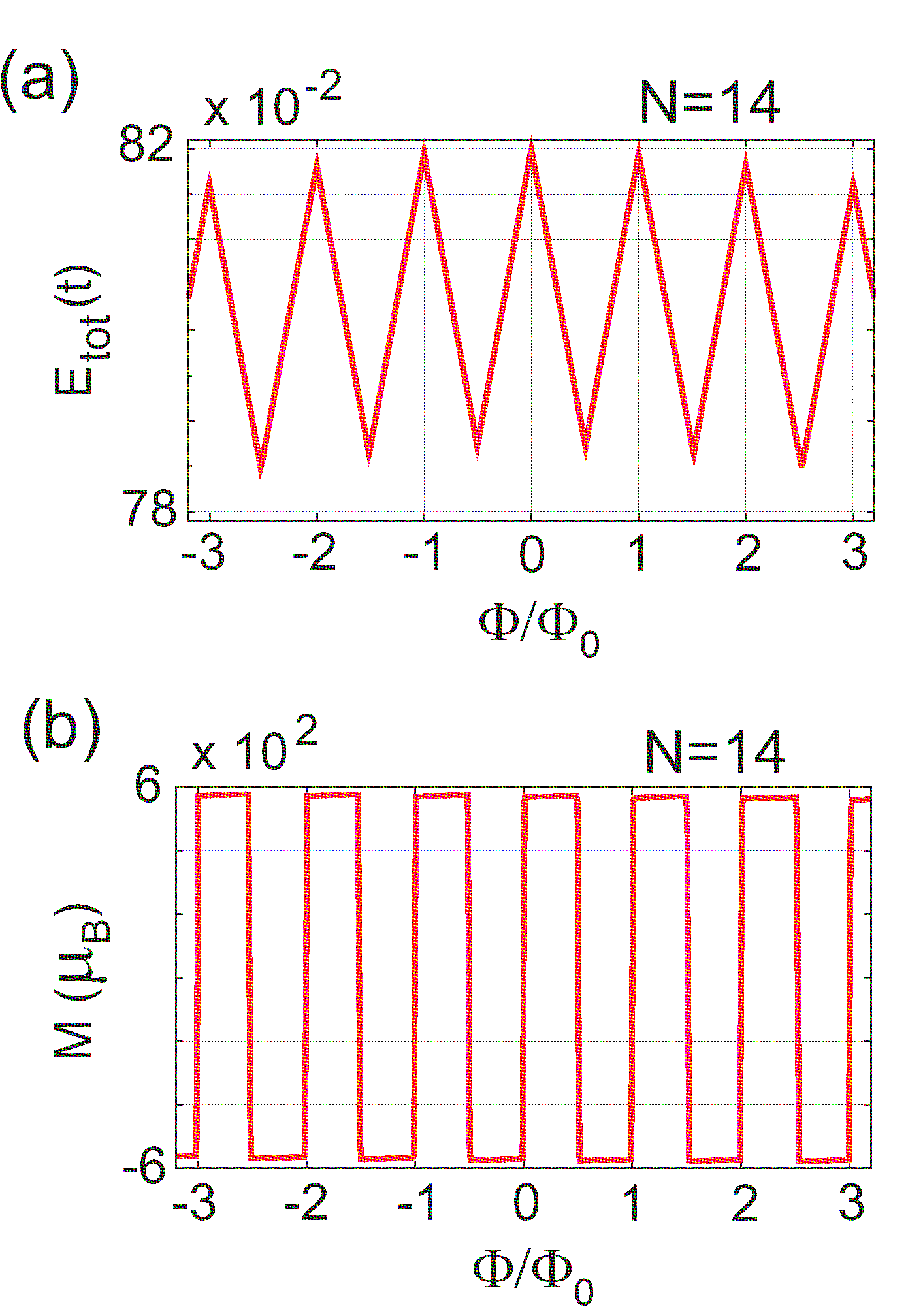}
\caption{
(Color online) 
   (a) TB total energies (sum over single-particle energies including spin) for $N=14$
       quasiparticles. 
   (b) Correposnding TB magnetization (in units of the Bohr magneton). 
       Energies in units of the TB hopping-parameter $t=2.7$ eV.
}

\label{tb_hex_arm2}
\end{figure}

\begin{figure}[t]
\centering\includegraphics[width=7.5cm]{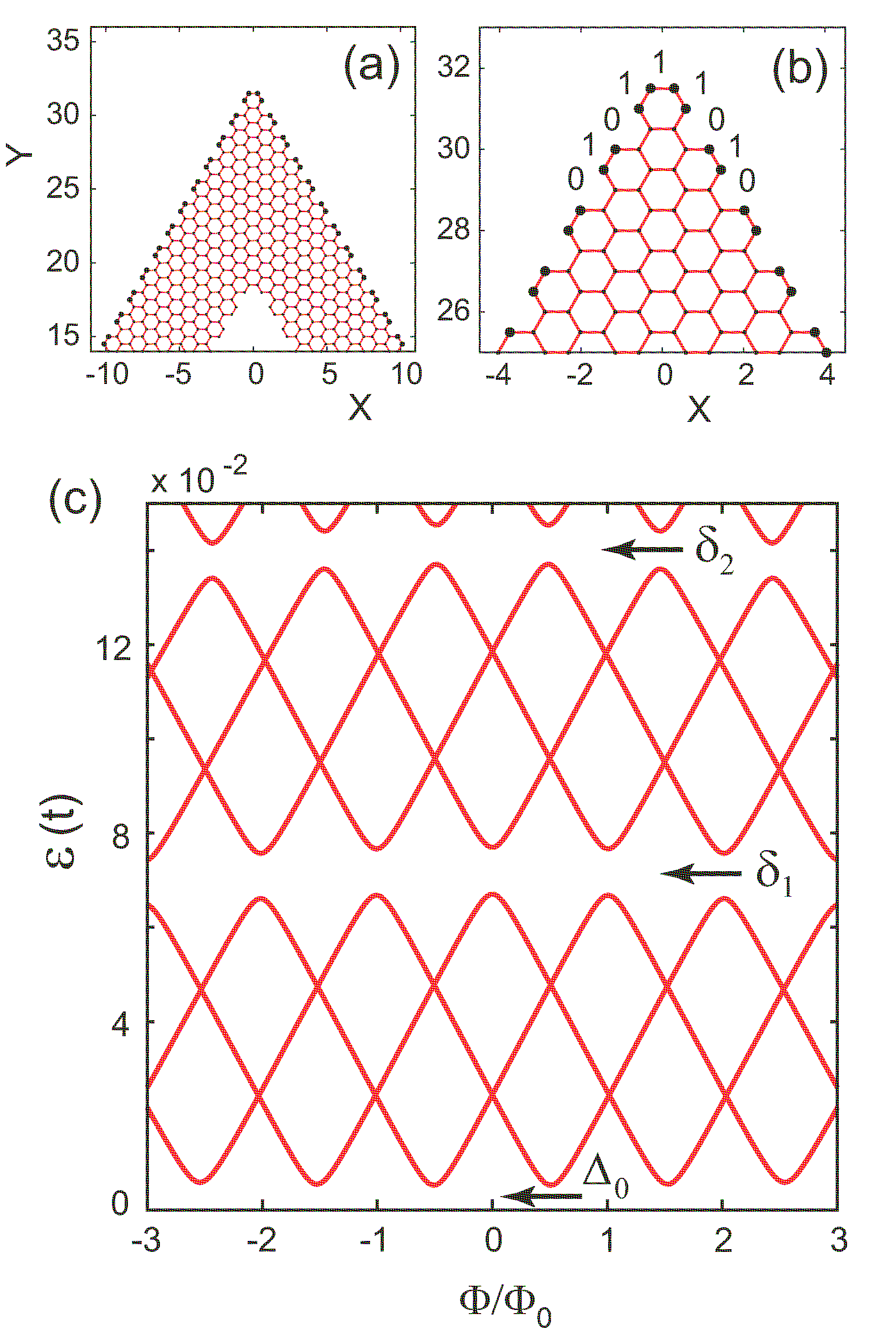}
\caption{
(Color online) 
   (a) Part of the trigonal graphene ring (2142 carbon atoms) with {\it armchair\/} edges.
   (b) Magnification of the corner of the trigonal ring shown in (a). 
       1 and 0 denote the short and long carbon dimers, respectively.
       Lengths in units of the graphene lattice constant $a_0=0.246$ nm.   
   (c) TB single-particle spectrum as a function of the magnetic flux (magnetic field).
       Energies in units of the TB hopping-parameter $t=2.7$ eV. The two arrows denoted $\delta_1$ and
       $\delta_2$ highlight band gaps. The arrow denoted by $\Delta_0$ indicates the opening of a
       gap at the Fermi level ($\varepsilon=0$) associated with generation of a rest mass ${\cal M}$.
       Note that the $\Delta_0$ gap is absent in the TB spectrum of the hexagonal graphene ring in 
       Fig.\ \ref{tb_hex_arm1}(b).
}
\label{tb_tri_arm}
\end{figure}

\begin{figure}[t]
\centering\includegraphics[width=6.5cm]{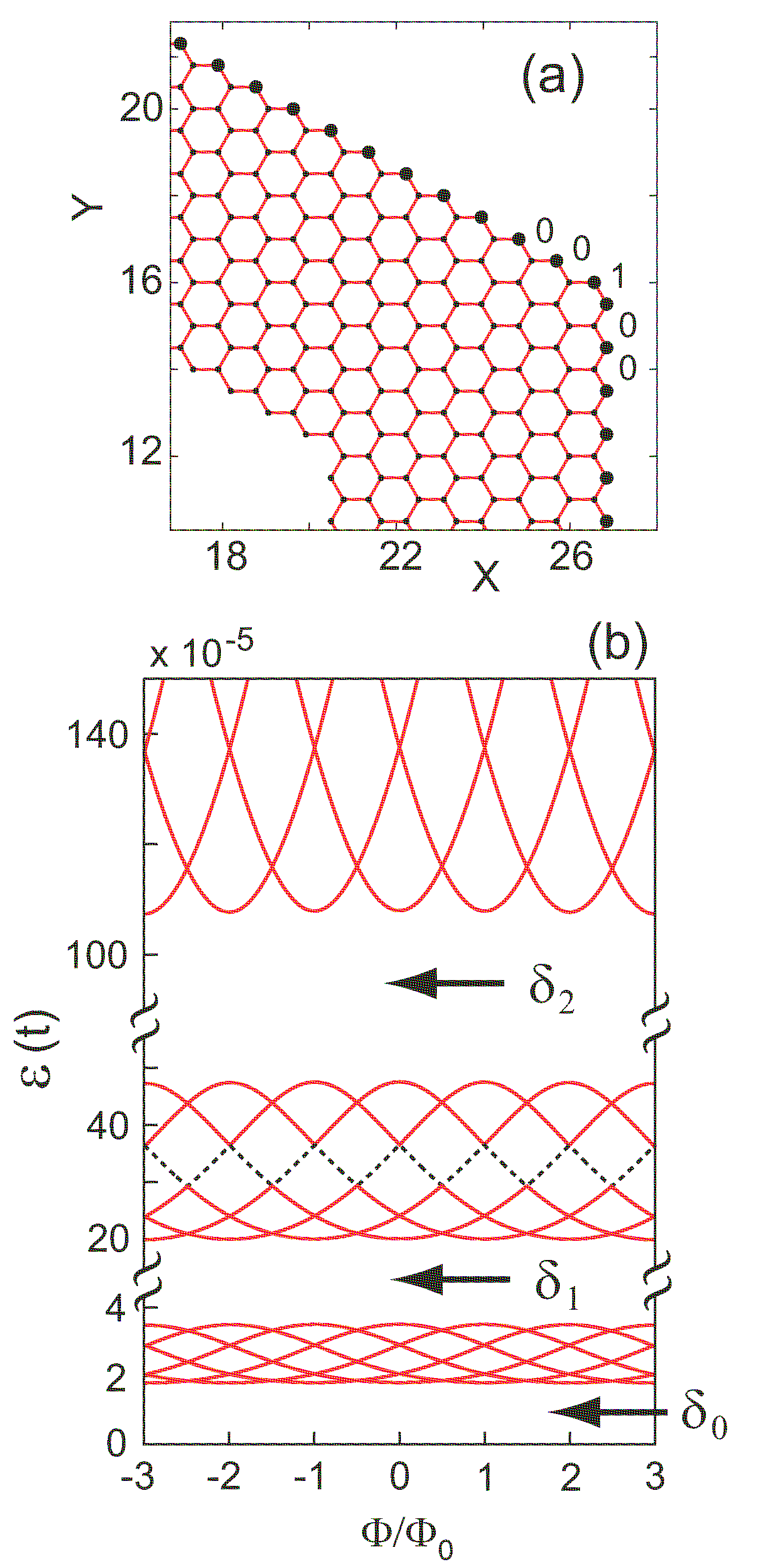}
\caption{
(Color online) 
   (a) Part of the hexagonal graphene ring (2688 carbon atoms) with {\it zigzag\/} edges.
       1 and 0 denote the short and long carbon dimers, respectively.
       Lengths in units of the graphene lattice constant $a_0=0.246$ nm.   
   (b) TB single-particle spectrum as a function of the magnetic flux (magnetic field).
       Energies in units of the TB hopping-parameter $t=2.7$ eV. The dashed black line denotes the
       Fermi level for $N=20$ quasiparticles. The arrows highlight the band gaps. 
}
\label{tb_hex_zz1}
\end{figure}

\begin{figure}[t]
\centering\includegraphics[width=5.5cm]{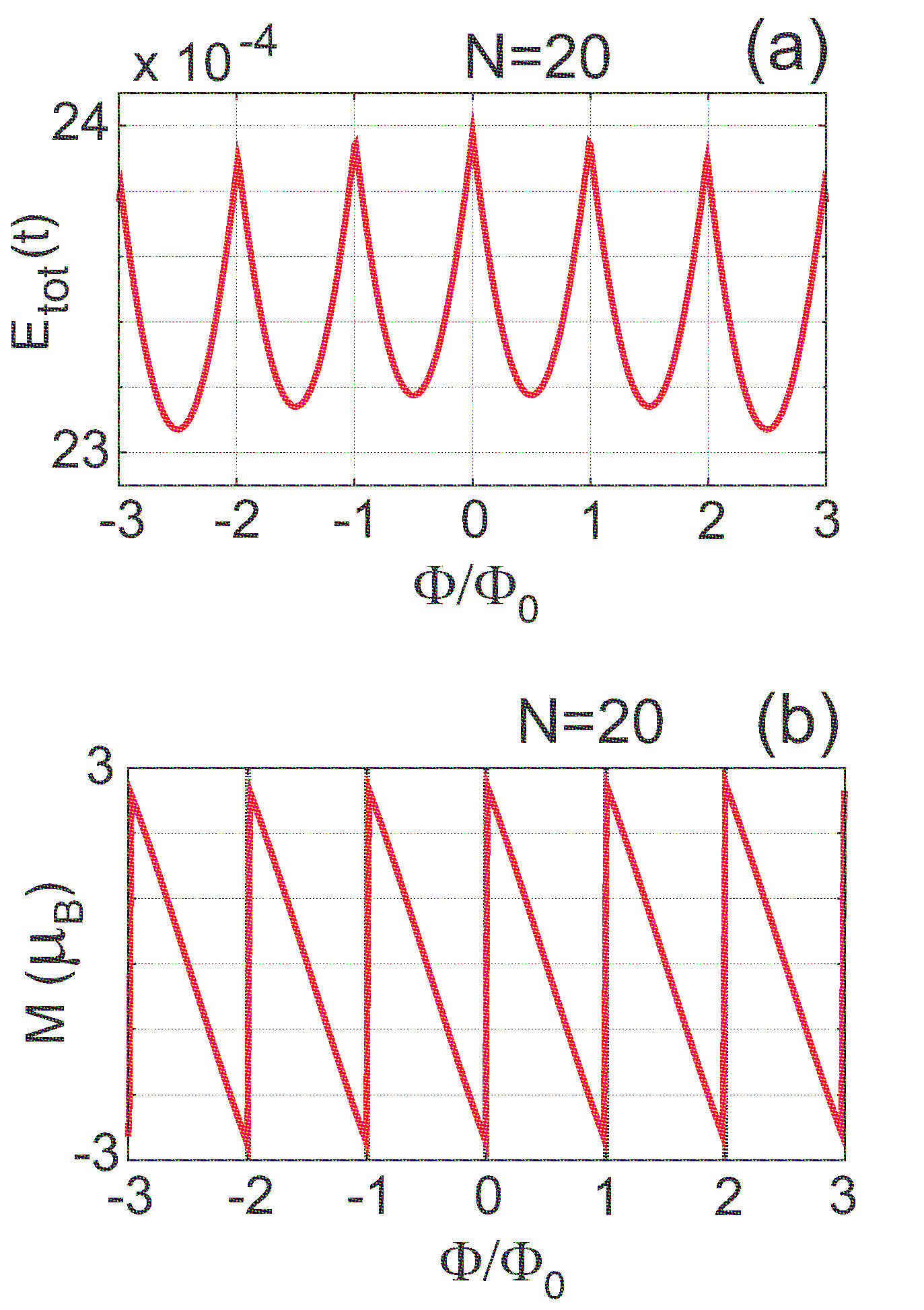}
\caption{
(Color online) 
   (a) TB total energies (sum over single-particle energies including spin) for $N=20$
       quasiparticles. 
   (b) Correposnding TB magnetization (in units of the Bohr magneton). 
       Energies in units of the TB hopping-parameter $t=2.7$ eV.
}
\label{tb_hex_zz2}
\end{figure}

\section{Tight-binding calculations}
\label{sectb}

In this section, we will describe the TB spectra and corresponding AB magnetizations (as a function of 
the magnetic flux) for three characteristic cases of planar graphene rings, and specifically for a 
{\it hexagonal armchair\/} ring, a {\it trigonal armchair\/} ring, and a {\it hexagonal zigzag\/} ring (all
three of similar dimensions). We note that the arms of the armchair rings studied here correspond to the 
class of perfect armchair nanoribbons referred to as metallic; 
\cite{tbgnr}$^{\text{(b),}}$\cite{dftgnr,dwgnr,revgnr} they exhibit a vanishing energy gap, $\Delta_0=0$, 
between the valence and conduction bands in tight-binding and continuum DW calculations. However, any 
perturbation (including the incorporation into a nanoring structure) may result\cite{tbgnr}$^{\text{(b)}}$ 
in the opening of a gap $\Delta_0>0$. The metallic aGNRs have a width corresponding to 
${\cal N}_W=3l-1$, $l=1,2,3,\ldots$ carbon atoms. Counting along a zigzag line in the middle of the arm 
(away from the corners), the hexagonal and trigonal armchair rings studied in this paper\cite{note5} have 
${\cal N}_W=14$.

To determine the single-particle spectrum [the energy levels $\varepsilon_i(B)$] in the 
tight-binding calculations for the graphene nanorings, we use the hamiltonian
\begin{equation}
H_{\text{TB}}= - \sum_{<i,j>} \tilde{t}_{ij} c^\dagger_i c_j + h.c.,
\label{htb}
\end{equation}
with $< >$ indicating summation over the nearest-neighbor sites $i,j$. The hopping
matrix element 
\begin{equation}
\tilde{t}_{ij}=t_{ij} \exp \left( \frac{ie}{\hbar c}  \int_{{\bf r}_i}^{{\bf r}_j} 
d{\bf s} \cdot {\bf A} ({\bf r}) \right), 
\label{tpei}
\end{equation}
where ${\bf r}_i$ and ${\bf r}_j$ are the positions of the carbon atoms
$i$ and $j$, respectively, and  ${\bf A}$ is the vector potential associated with the
applied constant magnetic field $B$ applied perpendicular to the plane of the nanoring.

The AB magnetization of the graphene ring is given by
\begin{equation}
M (\Phi) = - S \frac{d {E_\text{tot}}}{d \Phi},
\label{pc}
\end{equation}
where the total energy 
\begin{equation}
E_{\text{tot}} (\Phi)  = \sum_{i,\sigma}^{\text{occ}} \varepsilon_i(\Phi)
\label{etot}
\end{equation}
is given by the sum over all occupied single-particle energies; the index $\sigma$ runs over spins. 
$\Phi=B S$ is the magnetic flux through the area $S$ of the graphene ring and $\Phi_0=h c/e$ is the flux 
quantum.

The diagonalization of the TB hamiltonian [Eq.\ (\ref{htb})] is implemented with the use of the 
sparse-matrix solver ARPACK. \cite{arpack} In calculating $E_{\text{tot}}$ [see Eq.\ (\ref{etot})], only the 
single-particle TB energies with $\varepsilon_i(B) > 0$ are considered. \cite{rech07}

The TB results exhibit significant differences between the three cases studied here. These differences 
fall into two categories, namely, (A) same edge termination but different shape and (B) same shape but 
different edge termination.

\subsection{Hexagonal versus trigonal armchair rings}
 
The shape of the hexagonal graphene ring with armchair edge terminations considered here, as well as the 
corresponding TB results regarding the single-particle spectrum, the total energy, and the 
magnetization are displayed in Fig.\ \ref{tb_hex_arm1} and Fig.\ \ref{tb_hex_arm2}, as a function of 
$\Phi/\Phi_0$. The shape of the trigonal graphene ring with armchair edge terminations considered 
here, as well as the corresponding TB results regarding the single-particle spectrum are displayed in
Fig.\ \ref{tb_tri_arm}. 

Both TB energy spectra in Fig.\ \ref{tb_hex_arm1}(b) (armchair hexagon) and Fig.\ \ref{tb_tri_arm}(c) 
(armchair triangle) are organized in braid bands separated by energy gaps. They exhibit, however, two main 
differences. The first concerns the composition of the braid bands, with the hexagonal ring exhibiting 
six-membered bands while the trigonal ring having three-membered bands. The sixfold and threefold groupings
are a reflection of the $Z_6$ and $Z_3$ point-group symmetry of the hexagonal and trigonal shapes,
respectively; these symmetries are fully taken into account by the DKP modeling in Sec.\ \ref{secdkp}.

The second important difference between the TB energy spectra in Fig.\ \ref{tb_hex_arm1}(b) and Fig.\ 
\ref{tb_tri_arm}(c) concerns the number and nature of energy gaps. Specifically, two regular superlattice 
gaps $\delta_1$ and $\delta_2$ are present in both cases; note the similar energy scale [a magnification of
the region around the $\delta_1$ gap is shown in Fig.\ \ref{tb_hex_arm1}(c)]. However, while a mass gap 
$\Delta_0$ is well developed for the trigonal ring [Fig.\ \ref{tb_tri_arm}(c)], no corresponding $\Delta_0$
gap is present in the spectrum of the hexagonal ring, where the $\varepsilon=0$ horizontal axis dissects
(splits in half) the corresponding sixfold braid band [Fig.\ \ref{tb_hex_arm1}(b)]. As we will show in Sec.\
\ref{secintdkp}, the gap $\Delta_0$, for the case of the armchair triangle, is consistent with the physics 
of a massive (but still relativistic) Dirac fermion, while the dissecting of the $\varepsilon=0$ sixfold 
band, in the case of the armchair hexagon, is consistent with the formation of a fermionic soliton 
\cite{jr76} built on a scalar $Z_2$ kink soliton (precisely, a train of six fermionic solitons attached to 
successive $Z_2$ kink/antikink solitons; see also Sec.\ \ref{seclag}).

\subsection{Armchair versus zigzag hexagonal rings}

As aforementioned, the TB results exhibit also significant differences between the cases 
of hexagonal rings with armchair and zigzag terminations. One such difference concerns the $B$-dependence of
the single-particle energies $\varepsilon(\Phi)$, which is piecewise linear for the armchair case 
[Fig.\ \ref{tb_hex_arm1}(b)], but piecewise parabolic for the zigzag case [Fig.\ \ref{tb_hex_zz1}(b)]; this 
maintains also in the total energies $E_{\text{tot}}(\Phi)$ [Fig.\ \ref{tb_hex_arm2}(a) and Fig.\ 
\ref{tb_hex_zz2}(a)]. For the AB magnetizations [see Eq.\ (\ref{pc})], this 
results in a characteristically different profile for the AB oscillations of $M (\Phi)$: step-staggeredlike
in the armchair case [Fig.\ \ref{tb_hex_arm2}(b)] and sawtoothlike [Fig.\ \ref{tb_hex_zz2}(b)] in the zigzag
case. The paraboliclike $B$-dependence in the zigzag edge case \cite{roma12.2} is reminiscent of the 
spectra of a {\it nonrelativistic\/} ideal metal ring. \cite{gefe88} In contrast (see below), the linear 
$B$-dependence (in conjunction with the other features of the hexagonal armchair spectrum) can be 
associated with the fully relativistic regime of Dirac fermions with position-dependent masses. 
\cite{jack12,jack81}

Both the armchair and zigzag single-particle spectra for hexagonal are organized in six-member braid bands 
separated by energy gaps. This sixfold grouping is a reflection of the $Z_6$ point-group symmetry of the 
hexagonal rings. The energy gaps in the zigzag case are comparable to the width of the braid bands, and 
both the gaps and the widths of the bands increase with higher energy [see Fig.\ \ref{tb_hex_zz1}(b)]; this 
is consistent with a nonrelativistic Kronig-Penney model. \cite{roma12} In Fig.\ \ref{tb_hex_zz1}(b), there 
are three energy gaps labeled as $\delta_0$, $\delta_1$, and $\delta_2$. Unlike the relativistic regime, in
the nonrelativistic limit a gap around $\varepsilon=0$ is unrelated to the particle mass, and for this 
reason we use the symbol $\delta_0$ (with a lower-case $\delta$) instead of $\Delta_0$ as was the case in 
Fig.\ \ref{tb_tri_arm}(c). We note that in the nonrelativistic regime the effective mass of the 
quasiparticle excitations is proportional to the inverse of the second derivative of the (approximately) 
parabolic spectra; see, e.g., Ref.\ \onlinecite{kive83}. 

\begin{figure*}[t]
\centering\includegraphics[width=14cm]{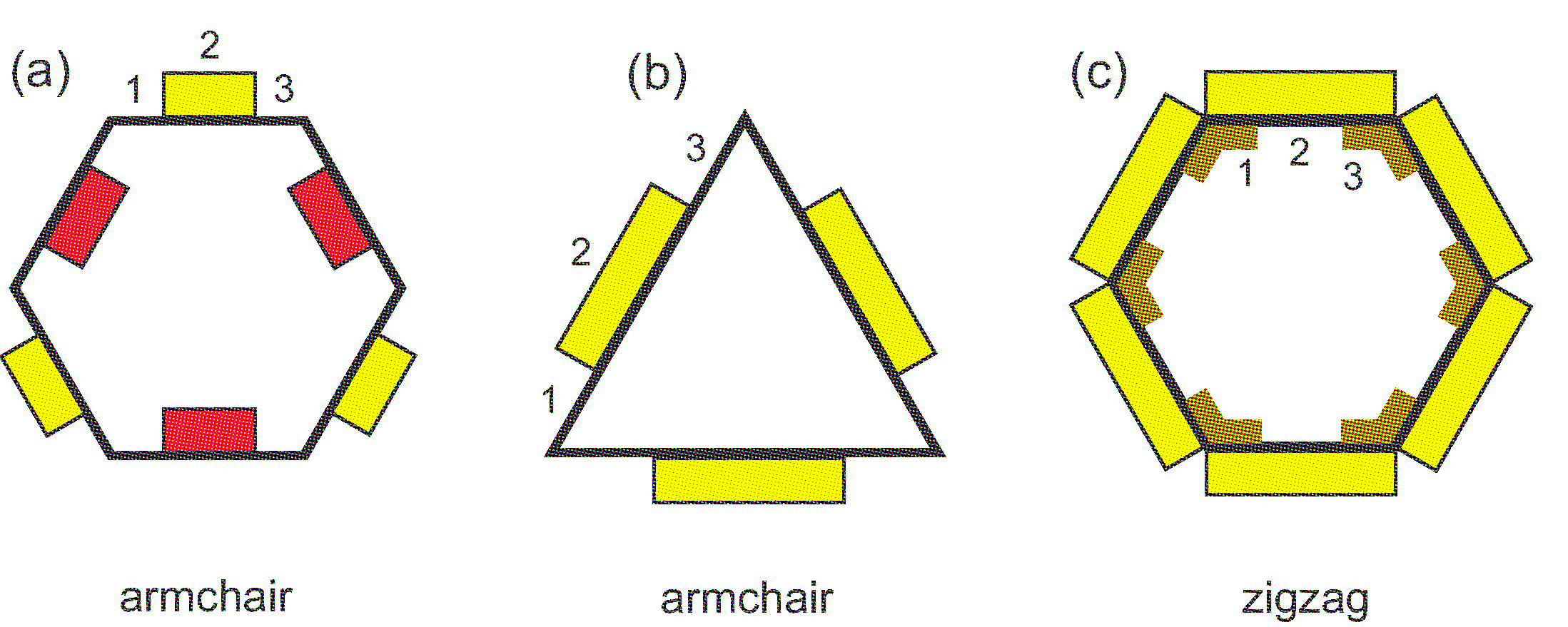}
\caption{
(Color online) 
Schematic representation of the $(V_i^{(n)}, m_i^{(n)})$, $i=1,\ldots,3$ and $n=1,\ldots,N_s$ 
square-step parameters entering in the DKP calculations: (a) in Sec.\ \ref{secdkphex}. (b) in Sec.\ 
\ref{secdkptri}. (c) in Sec.\ \ref{secdkphexzz}. Each side of the polygon is divided in three lengths 
$L_1$, $L_2$, and $L_3$. Boxes outside a polygon (online yellow) represent values $m_0>0$. Boxes inside a 
polygon (online red) represent values $-m_0 <0$. The nonzero values for the $V_1^{(n)}$ and $V_3^{(n)}$
in (c) are portrayed by thick lines (online brown) in the interior of the schematic hexagon. Zero values of 
parameters are not highlighted.     
}
\label{DKP_sch_para}
\end{figure*}

\section{Dirac-Kronig-Penney superlattice}
\label{secdkp}

As was shown for a semiconductor ring using a nonrelativistic superlattice approach, the AB single-particle
spectrum exhibits energy gaps demarcating Bloch bands when a scatterer is placed on the ring. \cite{imry83}
In this context, the energy gaps that appear in Fig.\ \ref{tb_hex_arm1}(b), Fig.\ \ref{tb_tri_arm}(c), and 
Fig.\ \ref{tb_hex_zz1}(b) indicate that the AB effect in polygonal graphene rings should be analyzed and 
modeled with the help of 1D Kronig-Penney-type superlattices, with the corners of the polygons providing a 
generalized analog to the ``scatterers'' of Ref.\ \onlinecite{imry83}. Specifically, we consider a 1D 
{\it relativistic\/} Dirac-Kronig-Penney model with unit cells built out of square potential barriers 
(developed in Ref.\ \onlinecite{mcke87} in the context of the physics of quarks).

The DKP model considered here is based on the 1D generalized Dirac equation, which has the form:
\begin{equation}
[E-V(x)] I \Psi + i \hbar v_F \alpha \frac{\partial \Psi}{\partial x} - \beta \phi(x) \Psi=0,
\label{direq}
\end{equation} 
with $v_F$ being the Fermi velocity of graphene, which replaces the speed of light $c$; 
$v_F/c \approx 1/300$. $V(x)$ is an electrostatic potential and $\phi(x)$ is a bosonic position-dependent 
scalar field. We note that Ref.\ \onlinecite{mcke87} uses $\phi(x) \equiv {\cal M} v_F^2 + V_s(x)$, with 
${\cal M}$ denoting the rest mass of a Dirac fermion (including the massless case) and the term $V_s(x)$ 
being referred to as the Lorentz scalar potential. Omitting the last term on the left of Eq.\ (\ref{direq})
reduces this equation to the massless Dirac-Weyl \cite{weyl29} one that underlies the majority of studies
in planar graphene.

The fermion field $\Psi$ is a twodimensional vector
\begin{equation}
\Psi = \left( \begin{array}{c} 
\psi_u \\
\psi_l \end{array} \right),
\label{psiferm}
\end{equation}
where the subscripts $u$ and $l$ stand for the upper and lower component, respectively. The $2 \times 2$
Dirac matrices $\alpha$ and $\beta$ can be \cite{nitt99} any two of the three Pauli matrices 
\begin{equation}
\sigma_1 = \left( \begin{array}{cc}
0 & 1 \\
1 & 0 \end{array} \right); \;
\sigma_2 = \left( \begin{array}{cc}
0 & -i \\
i &  0 \end{array} \right); \;
\sigma_3 = \left( \begin{array}{cc}
1 &  0 \\
0 & -1 \end{array} \right).
\label{paul}
\end{equation}
 
For example in the Dirac representation, one has $\alpha^D=\sigma_1$ and $\beta^D=\sigma_3$. $I$ is the
$2 \times 2$ identity matrix. We stress that
the energy spectra of the DKP model are independent of the specific representation used for $\alpha$ and
$\beta$. Below we will often use the notation $m(x)$, instead of $\phi(x)$, to stress the fact that 
$\phi(x)$ can be considered as a position-dependent mass term. 

In the DKP modeling of the TB results, we assign to the $n$th side ($n=1,\ldots,N_s$) of the polygonal ring
a number ($J$) of square potential steps (regions) denoted as $(V_i^{(n)}, m_i^{(n)})$, $i=1,\ldots,J$.
We note again that the electrostatic potentials $V_i^{(n)}$ enter the 1D Dirac equation [Eq.\ (\ref{direq}]
through the energy term $(E-V) I \Psi$, while the mass terms $m_i^{(n)}$ replace the scalar potential
in the term $\beta \phi(x) \Psi$; as a result these two potentials lead to different physical behavior in
the relativistic regime.

The building block of the DKP model is a 2$\times$2 wave-function matrix ${\bf \Omega}$ formed by the 
components of two independent 2$\times$1 spinor solutions of the onedimensional first-order Dirac 
equation. ${\bf \Omega}$ plays \cite{mcke87} the role of the Wronskian matrix ${\bf W}$ 
\cite{note1} used in the second-order nonrelativistic KP model; it is defined as follows at a point $x$ of the 
unit cell (here we use the Dirac representation): 
\begin{equation}
{\bf \Omega}_K (x) = \left( \begin{array}{cc}
e^{i K x} & e^{-i K x} \\
\Lambda e^{i K x}& -\Lambda e^{-i K x} \end{array} \right),
\label{ome}
\end{equation}
where 
\begin{equation}
K^2=\frac{(E-V)^2-m^2 v_F^4}{\hbar^2 v_F^2}, \;\;\; \Lambda= \frac{\hbar v_F K}{E-V+m v_F^2}.
\label{klam}
\end{equation}
We note again that, unlike the case of the original Dirac equation, \cite{dira28}  $m$ here is not a 
constant, but it may take different values from one region to the next. The transfer matrix for a given 
region (extending between two matching points $x_1$ and $x_2$) is the product 
${\bf M}_K (x_1,x_2)= {\bf \Omega}_K (x_2){\bf \Omega}_K^{-1} (x_1)$;  this latter matrix depends only on 
the width $x_2-x_1$ of the region, and not separately on $x_1$ or $x_2$. The relativistic ${\bf M}$ 
matrices defined here correspond to those considered \cite{gilmbook} in the case of a nonrelativistic 
superlattice in Ref.\ \onlinecite{roma12}.

The transfer matrix corresponding to the $n$th side of the hexagon is the product
\begin{equation}
{\bf t}_n = \prod_{i=1,J} {\bf M}_K (x_i,x_{i+1}),\;\;\; x_1=0,\; x_{J+1}=L,
\label{tside} 
\end{equation}
with $L$ being the (common) length on the hexagon side. The transfer matrix associated with the complete 
unit cell (around the ring) is the product
\begin{equation}
{\bf T}=\prod_{n=1}^{N_s} {\bf t}_n, 
\label{thex}
\end{equation}
where $N_s$ is the number of sides of the polygonal shape considered ($N_s=3$ and $N_s=6$ for a triangle
and hexagon, respectively).
 
Following  Ref.\ \onlinecite{imry83}, we consider the supperlattice generated from the virtual 
periodic translation of the unit cell as a result of the application of a magnetic field $B$ perpendicular 
to the ring. Then the AB energy spectra are given as solutions of the dispersion equation
\begin{equation}
\cos \left[ 2\pi(\Phi/\Phi_0+\eta) \right] = \Tr[{\bf T}(E)]/2,
\label{disrel} 
\end{equation}
where we have explicitly denoted the dependence of the r.h.s. on the energy $E$. The presence 
($\eta=1/2$) or absence ($\eta=0$) of an additional flux-shift in the relativistic or nonrelativistic 
case, respectively, follows through a comparison of the patterns of AB oscillations of a 
Dirac/Schr\"{o}dinger electron in the limiting case of an ideal metallic circular ring. 
\cite{cota07,gefe88}

\section{DKP interpretation of tight-binding case studies}
\label{secintdkp}

In this section, we demonstrate that our DKP modeling can capture the essential physics underlying the TB 
spectra in Fig.\ \ref{tb_hex_arm1}(b), Fig.\ \ref{tb_tri_arm}(c), and Fig.\ \ref{tb_hex_zz1}(b). In this 
respect, it also provides a framework for unifying the broad variety of behaviors of the TB spectra
of graphene nanorings. A schematic representation of the parameter sets used in our
DKP simulations is given in Fig.\ \ref{DKP_sch_para}.

\begin{figure*}[t]
\centering\includegraphics[width=16cm]{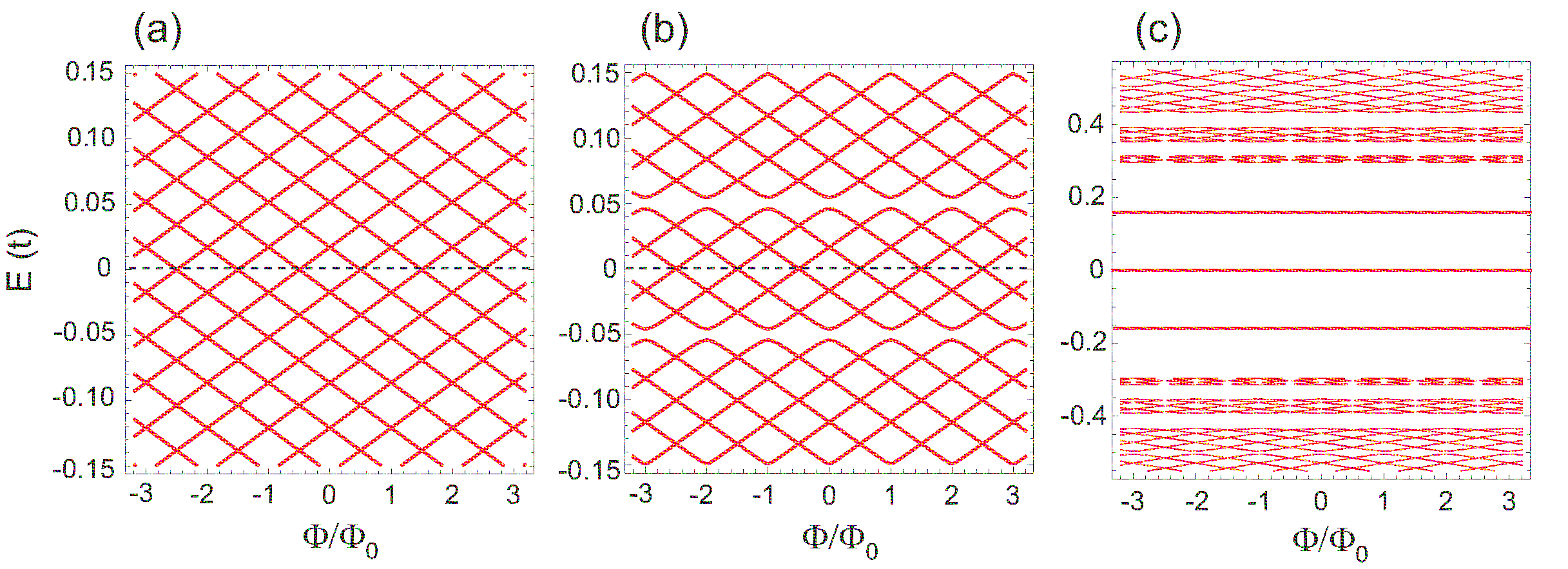}
\caption{
(Color online) 
Spectra from the DKP model (relativistic regime) corresponding to the schematic case (hexagon) in Fig.\ 
\ref{DKP_sch_para}(a); see text and Eqs.\ (\ref{l123}-\ref{m123}) for the full set of parameters employed.
(a) $m_0=0$ and any $V_i$. Note the absence of band gaps due to the Klein paradox.
(b) $m_0=0.01 t/v_F^2$. Note the similarity of the spectrum with that of the armchair graphene ring in 
Fig.\ \ref{tb_hex_arm1}(b). (c) $m_0=0.30 t/v_F^2$. The horizontal lines result from suppression of the AB 
oscillations due to strong localization. The dashed line indicates the energy zero.
}
\label{DKP_rel_hex}
\end{figure*}

\subsection{Armchair hexagonal ring and position-dependent-mass relativistic regime}
\label{secdkphex}

First we attempt a solution corresponding to the generalized Dirac equation (\ref{direq}) with $\phi(x)=0$.
In Fig.\ \ref{DKP_rel_hex}(a), we display the DKP spectra of a massless excitation (i.e., $m_i^{(n)}=0$ for 
all $i$ and $n=1,\ldots,6$). This massless DKP spectrum does not exhibit any gaps 
and it is strictly linear and periodic (with period $\Phi_0$) as a function of $\Phi$; this correlates with
the behavior of a free massless fermion, as in the case of 2D graphene. We note that this gapless spectrum
remains unchanged even when we consider in addition electrostatic potential steps, $V_i^{(n)} > 0$, a fact
that is a reflection of Klein tunneling. \cite{klei29,kats06}

However, the TB spectrum in Fig.\ \ref{tb_hex_arm1}(b) exhibits energy gaps (denoted as $\delta_1$ and
$\delta_2$, and highlighted by arrows), which require consideration of potential barriers in the DKP 
modeling. In the spirit of earlier investigations of real-space superlattices in 2D graphene, 
\cite{cohe08,peet08,fert10} we consider first a constant mass $m_i^{(n)}={\cal M}>0$, and alternating 
$V_i^{(n)}=V_0>0$ and $V_i^{(n)}=0$ steps in consecutive regions (see Sec.\ \ref{secdkp}). However, 
calculations with this choice show an opening of an energy gap 
at $\varepsilon=0$, a fact that conflicts with Fig.\ \ref{tb_hex_arm1}(b); in addition, it does not 
preserve the particle-hole symmetry of the TB spectra.

A crucial feature of the TB armchair spectra in Fig.\ \ref{tb_hex_arm1}(b) is the presence of zero-energy 
states (at half-integer fluxes). In order to capture this feature, and in light of our failed choices
(see above), we attempt next to use a non-vanishing position-dependent scalar field $\phi(x)$ [denoted
also as $m(x)$] in Eq.\ (\ref{direq}), and set the electrostatic potential $V(x)=0$. Recalling certain key
elements in the theory of {\it trans\/}-polyacetylene pertaining to zero-energy solitonic modes, 
\cite{jack12,jack81} we employ in our DKP transfer-matrix solution of Eq.\ (\ref{direq}) a scalar potential
$\phi(x)$ of the form $m(x)=-m(-x)$. Consequently, we divide each side of the hexagon in three parts ($J=3$)
of length 
\begin{equation}
L_1^{(n)}=a,\;\;L_2^{(n)}=b,\;\;L_3^{(n)}=a,
\label{l123}
\end{equation} 
and assign values 
\begin{equation}
V_1^{(n)}=V_2^{(n)}=V_3^{(n)}=0, 
\label{v123}
\end{equation}
and 
\begin{equation}
m_1^{(n)}=m_3^{(n)}=0,\;\;\;\;m_2^{(n)}=(-1)^n m_0. 
\label{m123}
\end{equation}
Note that the index $n=1, \ldots, 6$ here is numbering the sides of the hexagon, 
and thus overall the position-dependent mass term in our model is antisymmetric around each corner of the 
haxagonal ring. A schematic representation of the above parameters [Eq.\ (\ref{l123})$-$ Eq.\ (\ref{m123})]
is given in Fig.\ \ref{DKP_sch_para}(a).

Fig.\ \ref{DKP_rel_hex}(b) displays the DKP spectrum calculated with the dispersion equation (\ref{disrel}) 
using the above parameter set with $m_0=0.01t/v_F^2$, and $a=8a_0$, $b=15a_0$. \cite{note2} One observes 
that, in addition to the piecewise linear $B$-dependence, the DKP spectrum faithfully reproduces
the two other central features of the TB spectrum [Fig.\ \ref{tb_hex_arm1}(b)]: (i) the zero-energy states 
at half-integer values of $\Phi/\Phi_0$ and (ii) the opening at higher energies of energy gaps demarcating 
emerging sixfold braid bands. 

The behavior of each arm of the armchair hexagonal ring as a domain similar to the dimerized domains
of the {\it trans\/}-polyacetylene has a deeper physical reason, which can be revealed if one considers
each arm of the hexagon as a perturbed armchair graphene nanoribbon. Indeed analytic expressions for the
energy dispersion of the aGNRs have been recently derived [see Refs.\ \onlinecite{tbgnr}(b) and
\onlinecite{revgnr}]; they have the form
\begin{equation}
E(k) = \pm |t_1 e^{ika}+t_2 e^{-ikb}|,
\label{et1t2}
\end{equation}
with $k$ being the wave vector along the direction of the edge. $a=a_0/(2 \sqrt{3})$ and $b=a_0/\sqrt{3}$,
with $a_0=0.246$ nm being the lattice constant of graphene. 
$t_1=-2 t \cos[p \pi/({\cal N}_W+1)] + \delta t_1$, $p=1,2,\ldots,{\cal N}_W$ and $t_2=-t + \delta t_2$,
with $\delta t_1$, $\delta t_2$ denoting the perturbation away from a perfect aGNR. The dispersion equation
(\ref{et1t2}) is similar to the tight-binding one describing a onedimensional chain of carbon atoms with 
bonds (hopping matrix elements) of alternating strength $t_1$ and $t_2$ (Kekul\'{e} structure). The 
spectrum $E(k)$ exhibits a mass gap $\Delta_0=|t_1-t_2|$ at $k=\pi/(a+b)$. This behavior is analogous to 
that of the linear-chain lattice TB model for {\it trans\/}-polyacetylene; see Eq.\ (2.1) in Ref.\ 
\onlinecite{heeg88}. In particular, when $t_1 \neq t_2$, Eq.\ (\ref{et1t2}) describes a single dimerized
domain breaking the 1D reflectional symmetry; when $t_1 = t_2$ (metallic aGNR), it describes a symmetric 
chain of atoms, which preserves the reflectional symmetry. We note that the factor underlying the formation 
of dimerized domains in {\it trans\/}-polyacetylene is the Peierls instability incorporated in the 
Su-Schrieffer-Heeger model.\cite{heeg88,jack81} The corresponding factor in armchair graphene nanoribbons 
and rings is topological in nature, i.e., it is a reflection of the lattice distortions of graphene due to 
the edge termination and the shape.

Further insight can be gained through the observation that the armchair edge by itself reflects the carbon 
dimerization. Indeed it exhibits shorter dimers (denoted by 1) alternating with longer ones (denoted by 0); 
see Fig.\ \ref{tb_hex_arm1}(a). Thus each arm of the graphene ring corresponds to one of two equivalent 
domains $\ldots, 1,0,1,0, \ldots$ or $\ldots, 0,1,0,1, \ldots$. Following this notation and going around a 
given corner, one gets symbolically $\ldots, 0,1,0,\underline{1,1},0,1,0, \ldots$, i.e., each corner 
(denoted by an underline) acts as a domain wall separating two alternative domains. 

\begin{figure*}[t]
\centering\includegraphics[width=15cm]{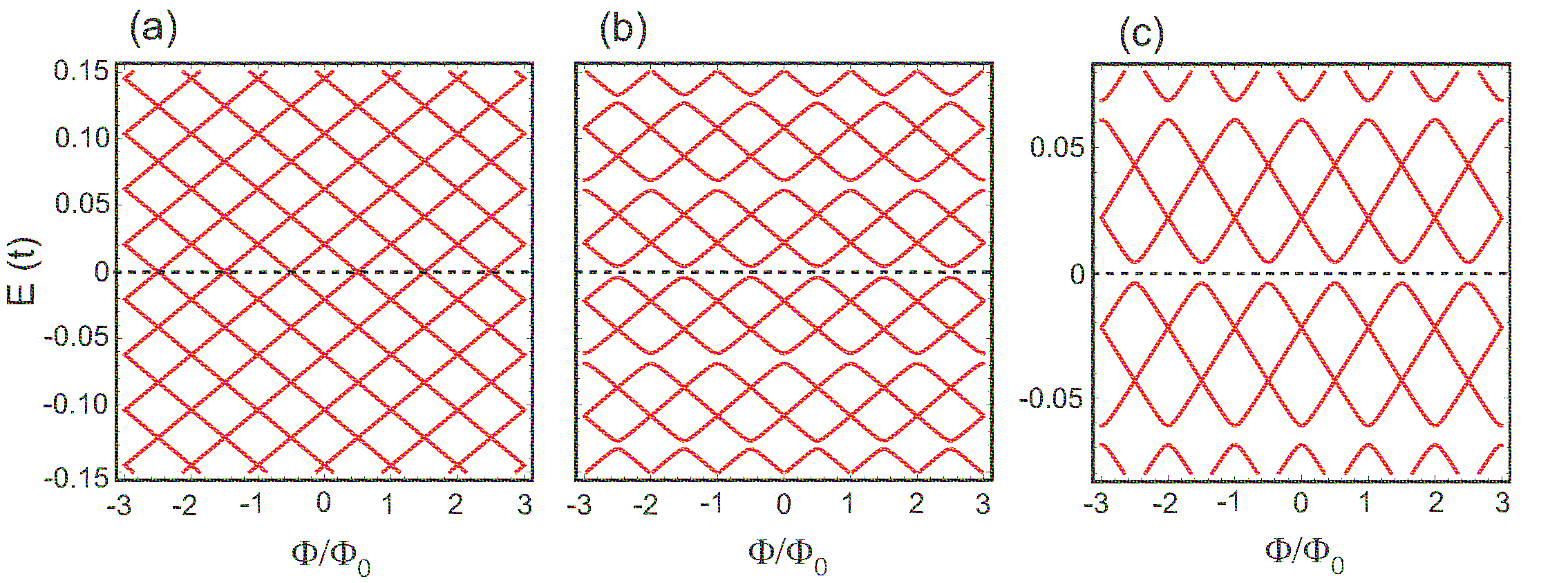}
\caption{
(Color online) 
Spectra from the DKP model (relativistic regime) corresponding to the schematic case (triangle) in Fig.\ 
\ref{DKP_sch_para}(b); see text and Eq.\ (\ref{m123tri}) for the full set of parameters employed. Panels (a) 
and (b,c) correspond to different choices of $m_0$ [denoted also as ${\cal M}$ here; see 
Eq.\ (\ref{m123tri})]. (a) $m_0=0$ and any $V_i$. Note the absence of band gaps due to the Klein paradox.
(b,c) $m_0=0.02 t/v_F^2$, and in (c) we display a magnification of the region around $E=0$. 
Note the similarity of the spectrum with that of the armchair graphene ring in
Fig.\ \ref{tb_tri_arm}(c). The dashed line indicates the energy zero.
}
\label{DKP_rel_tri}
\end{figure*}

In an {\it infinite\/} polyacetylene chain and at the position of the domain wall, a strongly localized 
{\it fermionic\/} soliton develops having a fractional charge $\pm 1/2$. \cite{jack12,jack81} However, the 
graphene nanoring is a {\it finite\/} system and overall it can carry only integer charges. 
\cite{raja82,kive82} Furthermore, the TB results, and their DKP analog in Fig.\ \ref{DKP_rel_hex}(b), 
indicate a moderate extent of localization at the corners, resulting from a non-negligible tunneling between
the corners. Strong solitonlike localization (exhibiting a 1/6 fractional charge at each corner) can be 
achieved for larger values of $m_0$. Indeed a large $m_0$ can localize a massless fermion, as is known from 
the (theoretical) trapping methodology in 2D graphene referred to as the infinite-mass boundary condition 
(introduced in Ref.\ \onlinecite{berr87} in the context of trapping neutrinos), as well as from the 
localization of massless quarks discussed in Ref.\ \onlinecite{mcke87}. The DKP spectrum for a large value 
$m_0 =0.30t/v_F^2$ is displayed in Fig.\ \ref{DKP_rel_hex}(c). It is seen that now the single-particle 
energies falling within the gap $-0.3 t < E < 0.3 t$ form horizontal straight lines because the 
corresponding AB oscillations have been suppressed due to vanishing of the tunneling between the fermionic
solitons at the corners; because of the localization no magnetic flux is trapped by the wave function on the
hexagonal ring. It is of interest to note that such a train configuration of fermionic solitons in 
an hexagonal ring may be referred to as a fractional Wigner crystallite. We note that besides the 
zero-energy fermionic soliton discussed in the context of the states of polyacetylene, \cite{jack12,jack81} 
Fig.\ \ref{DKP_rel_hex}(c) indicates the emergence of two polaroniclike states \cite{camp01,camp82} with 
energies $\approx \pm 0.16 t$ falling within the gap.

\subsection{Armchair trigonal ring and constant-mass relativistic regime}
\label{secdkptri}

The results of our DKP calculations [associated with Eq.\ (\ref{direq})] for the armchair trigonal ring are
shown in Fig.\ \ref{DKP_rel_tri}. 

Fig.\ \ref{DKP_rel_tri}(a) displays the DKP spectra of massless excitations 
(i.e., $m_i^{(n)}=0$ for all $i$ and $n=1,2,3$). As was the case with the hexagonal ring [Fig.\ 
\ref{DKP_rel_hex}(a)], the massless DKP spectrum in Fig.\ \ref{DKP_rel_tri}(a) does not exhibit any 
energy gaps and it is strictly linear and periodic (with period $\Phi_0$) as a function of $\Phi$ (or
equivalently the magnetic field $B$). We again note that this gapless spectrum remains unchanged even when 
we consider in addition electrostatic potential steps, $V_i^{(n)} > 0$, a fact that is a reflection of Klein
tunneling. \cite{klei29,kats06}

However, the TB spectrum in Fig.\ \ref{tb_tri_arm}(c) exhibits energy gaps (denoted as $\Delta_0$, 
$\delta_1$ and $\delta_2$, and highlighted by arrows), which require inclusion of potential barriers in
the DKP modeling. Following the analogy with the {\it trans\/}-polyacetylene, it is apparent that the
opening of the $\Delta_0$ gap indicates the absence of domain alternation. Namely, the armchair trigonal 
ring represents a realization of a single domain extending along the full length of the triangle. Thus the
corners of the triangle act a scatterers instead of domain walls as in the case of the hexagonal ring
(see Sec.\ \ref{secdkphex}). Indeed, following the notation of 1 and 0 introduced above for the short and 
long dimers [Fig.\ \ref{tb_hex_arm1}(a)], and going around the corner of the trigonal ring in 
Fig.\ \ref{tb_tri_arm}(a) [or Fig.\ \ref{tb_tri_arm}(b)], one gets the sequence 
$\ldots,0,0,0,{\underline {1,1,1}},0,0,0,\ldots$,
which is in agreement with the presence of the same domain on both sides of the $\pi/3$ corner.
Consequently, in the DKP modeling we keep the same parametrization as in Eqs.\ 
(\ref{l123}) and (\ref{v123}), but we replace Eq.\ (\ref{m123}) by
\begin{equation}
m_1^{(n)}=m_3^{(n)}=0,\;\;\;m_2^{(n)}=m_0={\cal M},\;\;n=1,2,3,
\label{m123tri}
\end{equation}
that is, the mass parameters are the same (not alternating) on the three arms of the trigonal ring [see
schematic representation in Fig.\ \ref{DKP_sch_para}(b)]; recall that ${\cal M}$ denotes the fermion
rest mass [see discussion below Eq.\ (\ref{direq})].

Fig.\ \ref{DKP_rel_tri}(b) [see also a magnification in Fig.\ \ref{DKP_rel_tri}(c)] displays the DKP spectrum 
calculated from the dispersion equation (\ref{disrel}) using the above parametrization [see Eqs.\  
(\ref{l123}), (\ref{v123}), and (\ref{m123tri})] with ${\cal M}=0.02t/v_F^2$, and $a=19a_0$, $b=10a_0$. 
\cite{note2} One sees that the DKP spectrum reproduces the two central features of the TB spectrum in Fig.\ 
\ref{tb_tri_arm}(c): (i) the gap $\Delta_0$ around $\varepsilon=0$ and (ii) the threefold braid bands which 
are separated by the gaps $\delta_1$ and $\delta_2$ at higher energies.

\begin{figure}[t]
\centering\includegraphics[width=6.0cm]{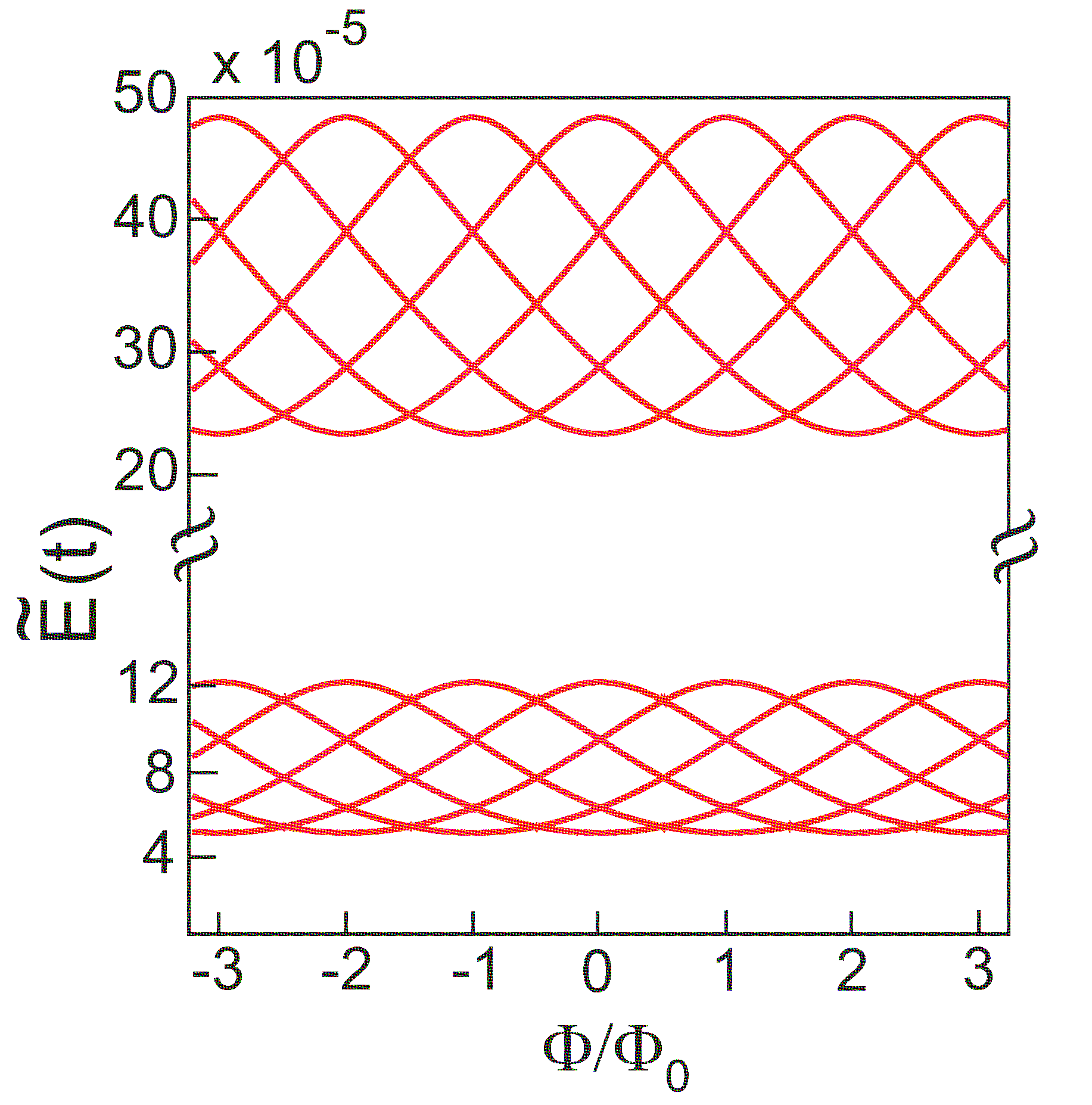}
\caption{
(Color online) 
Spectra from the DKP model ({\it nonrelativistic\/} regime) corresponding to the schematic case (hexagon) in 
Fig.\ \ref{DKP_sch_para}(c); see text and Eqs.\ (\ref{l123zz}-\ref{v123zz}) for the full set of parameters 
employed. The energies do not contain the 
rest-mass contribution, \cite{note3} that is, $\widetilde{E}(t)=E-{\cal M} v_F^2$, where $E$ is the
value obtained from solution of the DKP model and ${\cal M}$ is determined by fitting $\widetilde{E}$ to the
TB results [Fig.\ \ref{tb_hex_zz1}(b)]. The rest mass determined to yield the best fit shown here, is given by
${\cal M}=42.06 t/v_F^2$. This mass is heavier than that of the electron ($2.10 t/v_F^2$), suggesting an 
analogy with leptons.
}
\label{KP_nonrel_hex}
\end{figure}

\subsection{Zigzag hexagonal ring and constant-mass nonrelativistic limit}
\label{secdkphexzz}

As aforementioned, the TB spectrum of the hexagonal graphene ring with zigzag edges [Fig.\ 
\ref{tb_hex_zz1}(b)] shows trends associated with a large-mass fermion in the nonrelativistic regime:
(i) almost-perfect parabolic $B$-dependence and (ii) energy gaps $\delta_0$, $\delta_1$ and $\delta_2$ 
that are as large as the width of the braid bands. We remind that, for a massless excitation 
(ultrarelativistic limit), the spectra are strictly linear and exhibit no gaps [see Figs.\ 
\ref{DKP_rel_hex}(a) and \ref{DKP_rel_tri}(a)]; consequently small or moderate mass terms (relativistic 
regime) will result in smaller energy gaps compared to the width of the braid bands [see Fig.\
\ref{DKP_rel_hex}(b) and Figs.\ \ref{DKP_rel_tri}(b) and \ref{DKP_rel_tri}(c)]. 

The sharply different physics (nonrelativistic versus relativistic behavior) underlying the TB spectra
of the zigzag and armchair rings originates from the different edge topology. Following the notation of 1 
and 0 introduced above for the short and long dimers [Fig.\ \ref{tb_hex_arm1}(a)], and going around the 
$2\pi/3$ corner of the zigzag ring in Fig.\ \ref{tb_hex_zz1}(a), one further gets the sequence 
$\ldots,0,0,0,{\underline 1},0,0,0,\ldots$ This, in addition to the anticipation of a nonrelativistic
regime, suggests that the entire hexagonal zigzag ring constitutes a single domain with embedded (corner)
impurities (electrostatic potential scatterers). 

In light of the above, using a large rest mass ${\cal M}$ in Eq.\ (\ref{klam}) [nonrelativistic limit, see
Eqs.\ (4.27) and (4.28) in Ref.\ \onlinecite{mcke87}], we were able to reproduce in the DKP model [see 
Fig.\ \ref{KP_nonrel_hex}] the overall trends of the TB spectrum [Fig.\ \ref{tb_hex_zz1}(b)] for the zigzag 
ring. The DKP parameters used \cite{note2} were as follows,
\noindent 
Lengths:
\begin{equation}
L_1^{(n)}=L_3^{(n)}=a=1.5 a_0, \;\;L_2^{(n)}=b=28 a_0.
\label{l123zz}
\end{equation}
Rest mass: 
\begin{equation}
{\cal M}=m_1^{(n)}=m_2^{(n)}=m_3^{(n)}=42.06t/v_F^2.
\label{m123zz}
\end{equation}
Potential barrier step at each corner: 
\begin{equation}
V=V_1^{(n)}=V_3^{(n)}=80 \times 10^{-5} t,\; {\text{with}}\; V_2^{(n)}=0.
\label{v123zz}
\end{equation}
A schematic representation of these parameters is given in Fig.\ \ref{DKP_sch_para}(c).

In the evaluation of the nonrelativistic limit of the DKP model, one often sets $E={\cal M} v_F^2 + 
\widetilde{E}$ for the positive energies [see Eq.\ (4.27) in Ref.\ \onlinecite{mcke87}]. The quantity 
calculated from the DKP model in this limit is $E$, while in comparing with the TB results [Fig.\ 
\ref{tb_hex_zz1}(b)] we plot in Fig.\ \ref{KP_nonrel_hex} $\widetilde{E}$ vs. $\Phi$. The value of ${\cal M}$
is determined by finding the best fit to the TB results.\cite{note3} It is expected
that further improvement in the agreement between the TB and DKP approaches can be achieved by employing 
smooth (rather than square-shaped) profiles for the potential barriers. The constant mass, ${\cal M}$, 
found by us here is twenty times larger than the rest mass of the electron, indicating an analogy with 
electronlike leptons, \cite{leptbook} rather than the electron itself.

We note that the large rest mass ${\cal M}$ found in this section is unrelated to the energy gap $\delta_0$
[see Fig.\ \ref{tb_hex_zz1}(b)]; $({\cal M} v_F^2/\delta_0 \sim 10^5)$. Indeed in the nonrelativistic 
regime, the effective mass is related to the second derivative of the band spectra.\cite{kive83} Our 
findings concerning the zigzag graphene ring are in agreement with the analysis of Ref.\ \onlinecite{kive83} 
regarding the TB spectra of (infinite) symmetric polyacene, which is a single chain of fused benzene rings 
and can be considered as the thinnest possible zigzag graphene nanoribbon.

\section{The underlying relativistic quantum field Lagrangian}
\label{seclag}

In the above (see Sec.\ \ref{sectb}), we identified characteristic patterns of tight-binding spectra in 
planar graphene nanorings and we described their dependence on the edge termination (armchair versus zigzag)
and the ring shape (hexagonal versus trigonal). Subsequently, we presented a unified interpretation of the TB
results using a Dirac-Kronig-Penney model (Sec.\ \ref{secdkp}), built upon the generalized 1D Dirac equation
[Eq.\ (\ref{direq})]. A central finding of our relativistic DKP analysis were the close analogies (found in 
Secs.\ \ref{secdkphex} and \ref{secdkptri}) between the behavior of armchair hexagonal and trigonal graphene 
nanorings and the physics of {\it trans\/}-polyacetylene.\cite{jack12,jack81,heeg88}  

A natural next step towards a deeper understanding of the connection of our findings to relativistic
quantum field theory is the elucidation of the underlying Lagrangian formalism. Motivated by the theory
of {\it trans\/}-polyacetylene,\cite{jack81} we write a total Lagrangian density
\begin{equation}
{\cal L} = {\cal L}_f + {\cal L}_\phi,
\label{lagr}
\end{equation}
which is the sum of (partial) Lagrangian densities for the fermion field $\Psi$ [Eq.\ (\ref{psiferm})] and  
the bosonic scalar field $\phi(x)$. We note that the scalar field $\phi(x)$ was denoted earlier also as 
$m(x)$ and was referred to as a position-dependent mass; see Sec.\ \ref{secdkp} and Sec.\ \ref{secdkphex}.

The stationary generalized Dirac equation [Eq.\ (\ref{direq})] can be derived from the fermionic Lagrangian 
density
\begin{equation}
{\cal L}_f = - i \hbar \Psi^\dagger \frac{\partial}{\partial t} \Psi
- i \hbar v_F \Psi^\dagger \alpha \frac{\partial}{\partial x} \Psi 
- \phi \Psi^\dagger \beta \Psi,
\label{lagrf}
\end{equation}
where we have neglected contributions from the electrostatic potentials (see discussion in Sec.\ 
\ref{secdkphex}).

In Eq.\ (\ref{lagrf}), the last term
\begin{equation}
{\cal L}_Y = - \phi \Psi^\dagger \beta \Psi,
\label{lagry}
\end{equation}
which depends on both the fermion $\Psi$ and scalar $\phi$ fields, has the form of a Yukawa coupling. 
${\cal L}_Y$ is the potential agent for rest-mass acquisition by the originally massless fermion [described 
by the first two terms in the right hand side of Eq.\ (\ref{lagrf})]. 

The Yukawa interaction is also used in the Standard Model \cite{glas61,wein67,sala68} to describe the 
coupling between the Higgs field and the massless quark and lepton fields (i.e., the fundamental fermion 
particles). Through spontaneous symmetry breaking \cite{note7,namb60,ande72,yl07} of the Higgs field 
[which is a complex $SU(2)$ doublet of four real scalar fields $\phi$], these fermions acquire a mass 
proportional to the vacuum expectation value of the Higgs field. \cite{note6,mcmabook,bedn08} 

The essential observation that we make here is that, although, due to the 1D character of the graphene rings,
the 1D Lagrangian in Eq.\ (\ref{lagr}) does not possess the full richness of the Lagrangian of the Higgs 
sector in the Standard Model, both share the central aspect of symmetry breaking and mass acquisition by a 
fermion via a Yukawatype interaction.

We turn next to the task of constructing the Lagrangian part ${\cal L}_\phi$ for the scalar field $\phi$,
which is of the general form\cite{rajabook,vachbook} (in $1+1$ dimensions, i.e., time plus one space 
dimension)
\begin{equation}
{\cal L}_\phi =  - \frac{1}{2} (\frac{\partial \phi}{\partial x})^2 - V(\phi),
\label{lagrphi}
\end{equation}
and which preserves the reflectional $Z_2$ symmetry, i.e., it is invariant under $\phi \rightarrow -\phi$. 
Note that we are interested in the adiabatic approximation, and thus we omit the time dependent terms in
${\cal L}_\phi$.

The emergence of a constant mass for the armchair trigonal ring (Sec.\ \ref{secdkptri}) could simply be 
accounted for by considering a constant value of $\phi=\phi_0={\cal M}$ in the fermion Lagrangian 
${\cal L}_f$; then one needs to pay no further consideration to the bosonic ${\cal L}_\phi$. However, a more
general position-dependent field, $\phi(x)=m(x)$, was found to be essential in our DKP-model analysis of the
armchair hexagonal ring (Sec.\ \ref{secdkphex}). In this case, $\phi(x)$ alternates between two unequal 
values $\pm \phi_0$, with $\phi_0=m_0$ [see Fig.\ \ref{DKP_sch_para}(a)]. This indicates breaking of the 
$Z_2$ symmetry of the solutions to the equation of motion derived from the Lagrangian in Eq.\ 
(\ref{lagrphi}). An expression for $V(\phi)$ which reproduces qualitatively the above behavior (including 
the trigonal ring case) is the socalled $\phi^4$, which corresponds to a quartic double well potential in 
$\phi$, i.e.,
\begin{equation}
V(\phi)=\frac{\xi}{4} (\phi^2 - \zeta^2)^2,
\label{vphi4}
\end{equation}
where $\xi$ and $\zeta$ are parameters.

With the potential in Eq.\ (\ref{vphi4}), the bosonic sector has the field equation (see Ch. 2.3 in Ref.\ 
\onlinecite{rajabook} and Ch. 1.1 in Ref.\ \onlinecite{vachbook})
\begin{equation}
-\frac{\partial^2 \phi}{\partial x^2} + \xi (\phi^2 - \zeta^2) \phi = 0.
\label{vphif} 
\end{equation}

Two solutions of Eq.\ (\ref{vphif}) are $\phi(x)=\pm \phi_0= \pm \zeta$; these solutions break the symmetry 
since $\phi_0 \neq 0$. Using these solutions in the Dirac Eq.\ (\ref{direq}), one obtains the standard 
constant-mass Dirac equation\cite{dira28} for the fermionic field. This case corresponds to the behavior of 
the armchair trigonal ring (Sec.\ \ref{secdkptri}), as well as to that of the zigzag hexagonal ring (Sec.\ 
\ref{secdkphexzz}).

In addition, however, Eq.\ (\ref{vphif}) has nonlinear solutions that interpolate between the locations
$\phi_0$ and $-\phi_0$ of the two minima of the $V(\phi)$ potential. One of these nonlinear solutions is 
called the $Z_2$ kink soliton and the other the $Z_2$ antikink soliton. The kink soliton is given by
\begin{equation}
\phi_k(x)= \zeta \tanh \left( \sqrt{\frac{\xi}{2}} \zeta x \right),
\label{ki}
\end{equation}
and the antikink soliton has the form 
\begin{equation}
\bar{\phi}_k(x) = - \phi_k(x).
\label{anki}
\end{equation}
We note that $\phi_k( \pm \infty)=\pm \zeta$ and $\bar{\phi}_k( \pm \infty)=\mp \zeta$, while 
$\phi_k(0)=\bar{\phi}_k(0)=0$. 

\begin{figure}[t]
\centering\includegraphics[width=6.0cm]{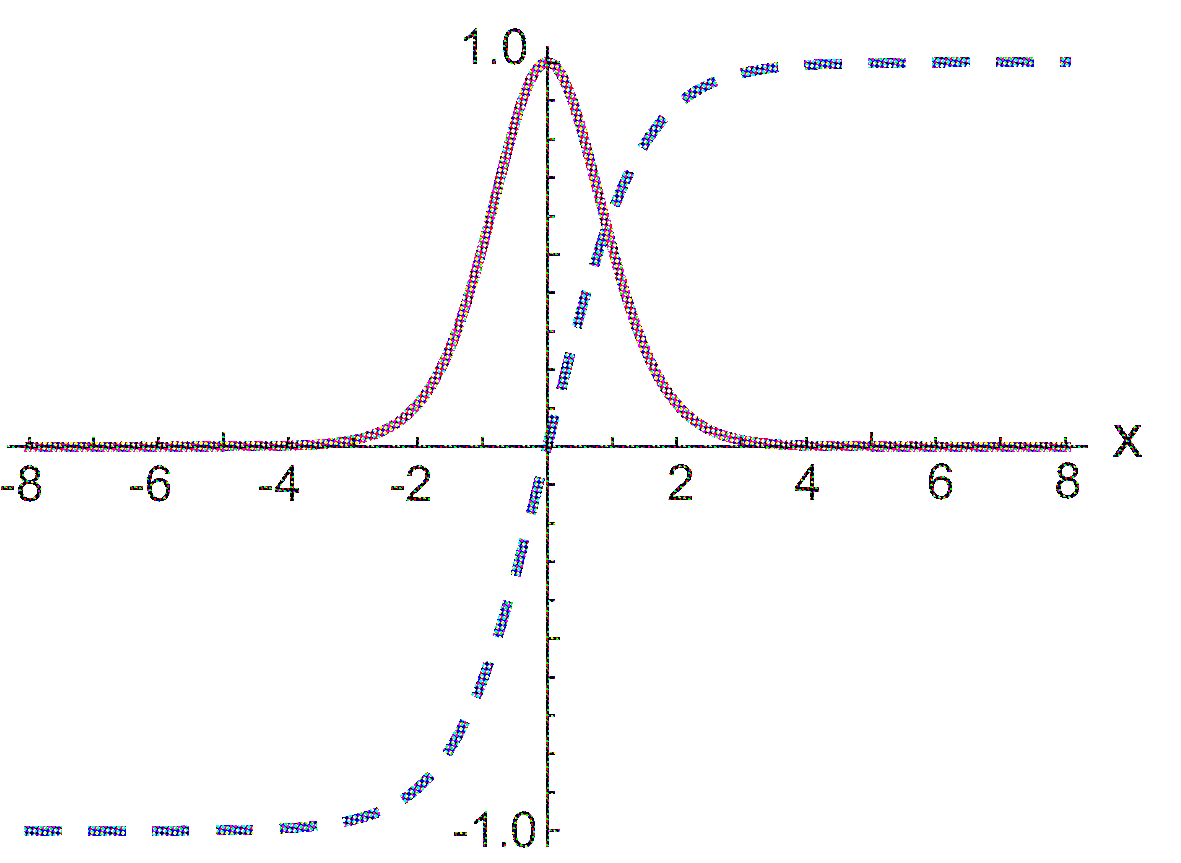}
\caption{
(Color online) 
Dashed line: The scalar $Z_2$ kink soliton [Eq.\ (\ref{ki})]. Solid line: The particle density (unnormalized)
of the corresponding enslaved fermionic soliton [Eq.\ (\ref{Psisol})]. The values $\xi=1$ and $\zeta=1$ were
used. The domain wall is located at $x=0$.
}
\label{Psiphisol}
\end{figure}

Using these kink or antikink scalar fields, the corresponding generalized Dirac equation [Eq.\ (\ref{direq})]
possesses {\it fermionic} soliton \cite{jack81} solutions of the form (here $\alpha=\sigma_2$, $\beta=\sigma_1$)
\begin{equation}
\Psi_S(x) \propto \left( \begin{array}{c}
 \exp \left(- \int_0^x \phi_k(x^\prime) dx^\prime \right)  \\
0 \end{array} \right).
\label{Psisol}
\end{equation}
The importance of these fermionic solitons lies in the fact that they have strictly {\it zero energies\/};
\cite{jack81} thus they fall into the particle/antiparticle (valence/conductance band) energy gap. 
Furthermore, they are localized at $x=0$, which is the domain wall between $x>0$ ($\phi_0$) and $x<0$ 
($-\phi_0$). The enslavement of the fermionic soliton $\Psi_S(x)$ by the scalar potential of the kink
soliton $\phi_k(x)$ is evident from Eq.\ (\ref{Psisol}). This is also reflected by the localization of
$\Psi_S(x)$ on the domain wall $x=0$ (see Fig.\ \ref{Psiphisol}).

It is apparent that such zero-mode fermionic solitons $\Psi_S(x)$ underlie qualitatively the behavior of the 
fermion excitations in the armchair hexagonal graphene nanoring (Sec.\ \ref{secdkphex}), with the corners of
the ring behaving as domain walls and the $m(x)$ stepwise function in the DKP model [$\phi(x)$ in Eq.\
(\ref{direq})] mimicking an alternation of $Z_2$ kink and antikink scalar solitons [Eqs.\ (\ref{ki}) and 
(\ref{anki})]; for a quantum-field-theory description of a train of alternating kinks and antikinks, see 
Ch.\ 1.7 in Ref.\ \onlinecite{vachbook}. 

\section{Conclusions} 
\label{seccon}

The paper investigated the different behavior of the Aharonov-Bohm spectra and magnetic-field induced 
oscillations for three characteristic cases of planar graphene nanorings, i.e., an hexagonal ring with 
armchair edge terminations, a trigonal ring with armchair edge terminations, and an hexagonal ring with 
zigzag edge terminations. The tight-binding results (Sec. \ref{sectb}) were analyzed with the help of a 1D 
relativistic Dirac-Kronig-Penney model \cite{mcke87} (Sec.\ \ref{secdkp}), which accounts for the virtual 
superlattice associated \cite{imry83} with the applied magnetic field. This analysis revealed unexpected 
topological effects and condensed-matter analogies with elementary particle physics. 

In particular, the behavior found by us for the armchair hexagonal ring (Sec.\ \ref{secdkphex}) is 
reminiscent of the extreme relativistic regime describing zero-energy fermionic solitons with fractional 
charge in quantum field theory\cite{jr76} and in the theory of {\it trans\/}-polyacetylene. 
\cite{jack12,jack81,heeg88,camp01} This regime results from a consideration 
of a modified (generalized) Dirac equation with a position-dependent mass 
term (or equivalently a position-dependent scalar bosonic field). In contrast, the quasiparticle excitations
in the armchair trigonal ring (Sec.\ \ref{secdkptri}) behave as relativistic Dirac fermions having a 
constant mass. A unification of these two dissimilar behaviors was presented in Sec.\ \ref{seclag} by 
introducing the underlying relativistic Lagrangian formalism for a fermionic and a scalar bosonic fields 
coupled via a Yukawa interaction. The Yukawa term in conjunction with the breaking of the $Z_2$ reflectional
symmetry of the scalar field may result in two outcomes, i.e., formation of a fermionic soliton (armchair 
hexagonal ring) or mass generation (armchair trigonal ring). 
The profoundly differing  behaviors found by us for the armchair hexagonal and trigonal rings (with both
sharing similar spatial dimensions), are  manifestations of the quantum topological nature of  this behavior,
as distinguished from ``quantum size effects'' which are length-scale dependent phenomena, originating from
spatial confinement of the electrons (quasiparticles, in general).\cite{xxx,yyy,zzz,www}  
  
The behavior of the zigzag hexagonal ring resembles the low-kinetic-energy {\it nonrelativistic\/} regime 
of a leptonlike fermion having a rest mass larger than that of the electron (Sec.\ \ref{secdkphexzz}).  
This behavior contrasts with the relativistic ones found for the aforementioned armchair rings, thus
highlighting the compounded topological and edge -termination effects. 

These findings\cite{note8,note9} highlight the potential of graphene nanosystems for providing a bridge between 
condensed-matter and particle physics, well beyond the paradigm of the massless neutrinolike fermion 
familiar from the 2D graphene sheet. Furthermore beyond the realm of graphene proper, where atomically 
precise narrow nanoribbons have already been synthesized,\cite{exgnr} we anticipate that our theoretical 
predictions could be tested using an ever expanding class of designer-Dirac-fermion manmade systems, 
such as optical lattices comprising ultracold atoms, \cite{java08,zhan12} or ``molecular'' 
\cite{mano12} and nanopatterned artificial graphene.\cite{pell09}

\begin{acknowledgments}
This work was supported by the Office of Basic Energy Sciences of the US D.O.E. under contract 
FG05-86ER45234. 
\end{acknowledgments}


\begin{thebibliography}{999}
\bibitem{note2}
The total length $L=2a+b$ equals approximately that of the side of the corresponding graphene ring.
\bibitem{note3}
In the nonrelativistic limit, for negative energies (representing holes with $-e$ charge), one sets 
$E=-{\cal M} v_F^2 + \widetilde{E}$ and uses a negative potential step $-V$. The resulting spectrum
is mirror-symmetric about the horizontal axis $\widetilde{E}=0$; note the we mentioned here holes
rather than antiparticles.
\bibitem{note7}
Symmetry breaking in elementary particle physics was introduced in Ref.\ \onlinecite{namb60}. For the
importance of symmetry breaking in bulk condensed-matter systems, see Ref.\ \onlinecite{ande72}. For 
symmetry breaking and the subsequent step of symmetry restoration in finite systems, see Ref.\ 
\onlinecite{yl07}.
\end{thebibliography}
\end{document}